\documentstyle[psfig]{nature_local}

\def\ltsima{$\; \buildrel < \over \sim \;$}
\def\simlt{\lower.5ex\hbox{\ltsima}}
\def\gtsima{$\; \buildrel > \over \sim \;$}
\def\simgt{\lower.5ex\hbox{\gtsima}}

\title{The discovery of a galaxy-wide superwind from a young massive galaxy at redshift $z \simeq 3$}

\author{R.J.~Wilman\affiliation{Department of Physics, University of Durham, South Road, Durham, DH1 3LE, UK},
  J.~Gerssen $^{*}$,
  R.G.~Bower $^{*}$,
  S.L.~Morris $^{*}$,
          R.~Bacon\affiliation{CRAL-Observatoire, 9 Avenue Charles-Andr\'e, 69230 Sain-Genis-Laval, France},
        P.T.~de Zeeuw\affiliation{Leiden Observatory, Niels Bohrweg 2, 2333 CA Leiden, The Netherlands} \&
        R.L.~Davies\affiliation{Astrophysics, University of Oxford, Keble Road, Oxford, OX1 3RH, UK}
}

\headertitle{}
\mainauthor{}

\summary{\noindent High-velocity galactic outflows, driven by intense bursts of star formation and black hole accretion, are invoked by current theories of galaxy formation$^{1}$ to terminate star formation in the most massive galaxies and to deposit heavy elements in the intergalactic medium. From existing observational evidence$^{2,3}$ on high-redshift galaxies, it is unclear whether such outflows are localized to regions of intense star formation just a few kiloparsecs in extent, or whether they instead have a significant impact on the entire galaxy and its surroundings. Here we present two-dimensional spectroscopy of a star-forming galaxy$^{4}$ at redshift z=3.09 (seen 11.5~gigayears ago, when the Universe was 20 per cent of its current age): its spatially extended Ly$\alpha$ line emission appears to be absorbed by HI in a foreground screen covering the entire galaxy, with a lateral extent of at least 100 kpc and remarkable velocity coherence. It was plausibly ejected from the galaxy during a starburst several 10$^{8}$ years earlier and has subsequently swept up gas from the surrounding intergalactic medium and cooled. This demonstrates the galaxy-wide impact of high-redshift superwinds.}\dates{6 December 2004}{5 May 2005}

\begin{document}
\maketitle

        The formation of galaxies requires gas to cool in haloes of dark matter that collapse under gravity from the expansion of the Universe. However, cooling alone overproduces bright galaxies at the present day, so models incorporate thermal conduction, photoionization and galaxy merging, together with additional `feedback'$^{1}$ in the form of galactic-scale outflows. The latter are powered by supernovae and massive stellar winds, or by relativistic winds and jets resulting from gas accretion onto supermassive black holes. Although starburst superwinds have been studied$^{5}$ in local dwarf galaxies (such as M82), observational evidence for their counterparts in young massive galaxies at high redshift has been less direct.

        In Lyman-break galaxies (LBGs) -- high-redshift galaxies with moderate masses and star-formation rates -- spectroscopic evidence points to powerful outflows$^{2,3}$ of interstellar gas, with absorption lines being blueshifted by several hundred kilometres per second from the galaxy's systemic velocity. The gravitationally lensed LBG MS1512-cB58 at redshift z=2.73 is a clear example$^{3}$: absorbing gas is outflowing at $\sim 255$~km~s$^{-1}$ at a rate exceeding the star-formation rate, and although it covers the entire star-forming region, this shows only that the radius of any shell exceeds $\sim 1$~kpc; it is not known whether the outflows are collimated and localized to star-forming regions, or whether they are galaxy-wide. The latter is suggested by the observed decrease in the intergalactic medium (IGM) H I opacity in background quasar sightlines that are close in projection to LBGs$^{6}$ (within 0.7~Mpc co-moving). This may indicate that LBGs drive superwinds at $\sim 600$~km~s$^{-1}$ for several 10$^{8}$~yr, or it might be a statistical fluctuation since there are few LBG-quasar pairs at the smallest separations.

        To characterize such outflows better via absorption studies, a background light source is needed with a spatial extent somewhat larger than an LBG stellar body or a quasar sightline. Such a source is provided by the recently discovered Ly$\alpha$-emitting blobs (hereafter LABs), associated with LBGs in the SSA22 protocluster at redshift z=3.09, a structure which is likely to evolve into a rich cluster of galaxies. With sizes $\sim 100$~kpc and Ly$\alpha$ luminosities $\sim 10^{44}$~erg~s$^{-1}$, the LABs resemble the emission-line haloes of high-redshift radio galaxies, but their radio emission is much fainter than in typical high-redshift radio galaxies$^{7}$. Models for LAB Ly$\alpha$ emission include superwind ejecta$^{8}$ and cooling radiation from gravitationally heated gas$^{9}$. Both LABs are submillimetre sources$^{10}$, with LAB-1 the more powerful and consistent with $\sim 1000$~M$_{\rm{\odot}}$~yr$^{-1}$ of star formation. LAB-2 has an X-ray detection, suggestive of an obscured active galactic nucleus$^{11}$, whereas LAB-1 does not.

        We recently observed the LAB Ly$\alpha$ haloes using integral-field spectroscopy, which, unlike conventional slit spectroscopy, gathers spatially resolved spectra over a two-dimensional area. Such information is essential for a complete picture of morphologically complex objects like the LABs. We used the SAURON Integral Field Spectrograph$^{12}$ on the William Herschel Telescope, providing moderate resolution spectroscopy ($\sim 4$\AA ~full-width at half-maximum, FWHM) over a $41 \times 31$~arcsec area, sampled with 0.95-arcsec lenslets. LAB-2 was observed for 15~h (in thirty 30-min sub-exposures), but unlike LAB-1$^{13}$ exhibits simple structure in Ly$\alpha$. At most positions the emission line is double-peaked, as shown in Fig. 1 and in the line profiles of individual regions (Figs. 2 and 3). The position of the profiles central minimum varies across the galaxy by less than 60~km~s$^{-1}$ in its rest-frame. The observing strategy of dithering between sub-exposures ensures that this feature is not a charge-coupled device (CCD) detector defect, nor is it due to a night-sky emission feature. The simplest interpretation is that part of the Ly$\alpha$ emission is scattered out of our line of sight by neutral hydrogen in a coherent foreground screen.

\begin{figure}[t]
\centerline{\psfig{file=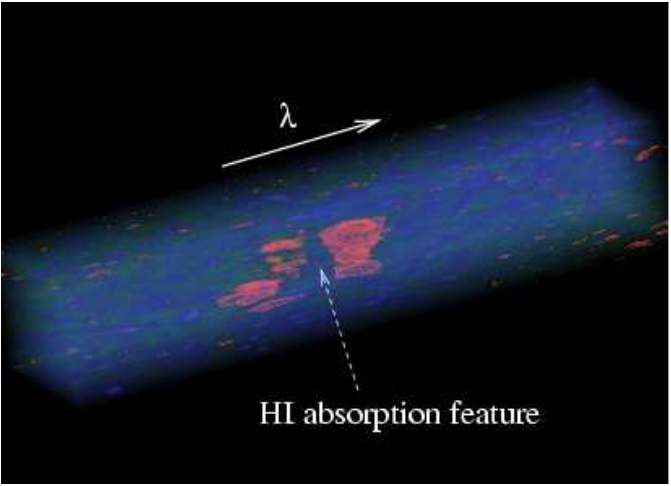,width=3.5in}}
\caption{\small Part of the data cube obtained from integral field spectroscopy of the z=3.09 galaxy LAB-2, showing the Ly$\alpha$ emission line. The absorption feature, interpreted as due to foreground HI, in indicated. A data cube is a three-dimensional structure for spectroscopic data from integral field spectrographs. Here the long axis corresponds to the dispersion axis and the other two axes represent spatial dimensions on the sky, that is, the data cube stores a spectrum for each point in a two-dimensional region of sky. With visualization software, the data-cube representation can provide useful qualitative insights into complex data sets. $\lambda$, wavelength.}
\end{figure}

        The line profiles were each modelled with a gaussian emission line and a superimposed Voigt profile absorber (all convolved with the instrumental response), as in Fig. 3. The HI column densities are N$_{\rm{HI}} \sim 10^{19}$~cm$^{-2}$, and the relative amplitudes of the red and blue line wings reflect velocity differences between the emission and absorption components. The absorber is typically blueshifted from the underlying peak by less than 250~km~s$^{-1}$ ; the emission shows a range in absolute velocity of $\sim 290$~km~s$^{-1}$, with individual linewidths of $\sim 1000$~km~s$^{-1}$ FWHM. At position B, no absorber is detected, perhaps because the underlying emission is more blueshifted than elsewhere.

\begin{figure}[t]
\centerline{\psfig{file=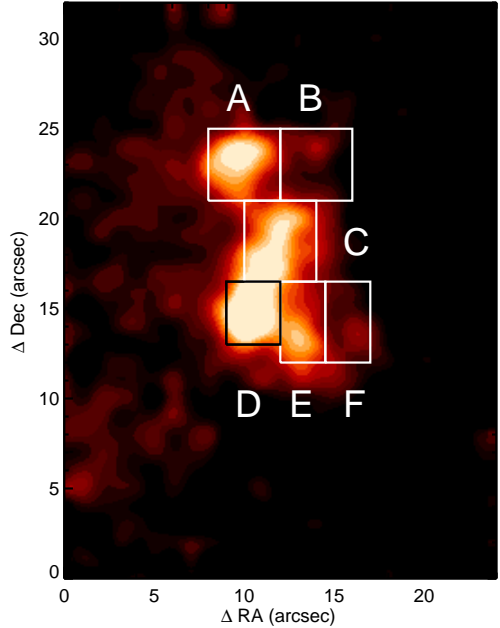,width=3.5in}}
\caption{\small A reconstructed image of the z=3.09 galaxy LAB-2 in the Ly$\alpha$ emission line. This shows the extent of its gaseous halo (10~arcsec on the sky equates to a physical distance of 76~kpc in this galaxy with the cosmological parameters H$_{\rm{0}}=70$~km~s$^{-1}$~Mpc$^{-1}$, $\Omega_{\rm{M}}=0.3$ and $\Omega_{\rm{\Lambda}}=0.7$). The image was derived from observations with the SAURON integral field spectrograph, by collapsing the data cube shown in Fig. 1 along the wavelength direction. The labels indicate regions for which one-dimensional Ly$\alpha$ emission line profiles have been extracted, as shown in Fig. 3. The underlying Ly$\alpha$ emission is likely to be from gas in the halo of the associated Lyman-break galaxy$^{4}$ (at position C), arising from a combination of superwind ejecta$^{8}$, cooling radiation$^{9}$, and gas photoionized by the obscured active nucleus$^{11}$.}
\end{figure}

        An absorbing shell is predicted to form when a starburst-heated hot gas bubble becomes over-pressurized relative to the interstellar medium and hence breaks out of the galactic disk, accelerates, and fragments as a result of Rayleigh-Taylor instabilities. Hot gas then escapes into the halo and forms a second shell of swept-up IGM. Evolutionary models for the Ly$\alpha$ emission from such superwinds$^{14}$ -- which are consistent with Hubble Space Telescope observations of local starburst galaxies$^{15}$ -- suggest that the LAB-2 absorber is in a late phase, where the shell has cooled and slowed sufficiently to absorb the underlying emission. Figure 4 illustrates the model.

        Because the absorbing shell's transverse extent is at least 100~kpc, it is probably a comparable distance in front of the Ly$\alpha$ source (although the constant absorber velocity implies that, if spherical, the radius must be considerably larger). At a velocity of 1000~km~s$^{-1}$, it takes 10$^{8}$~yr to travel 100 kpc. Although 1000~km~s$^{-1}$ is the expected level of the initial velocity, it has since slowed owing to mass loading from the IGM and work against the galaxy's gravity; this implies an age of several 10$^{8}$~yr.

\begin{figure}[t]
\centerline{\psfig{file=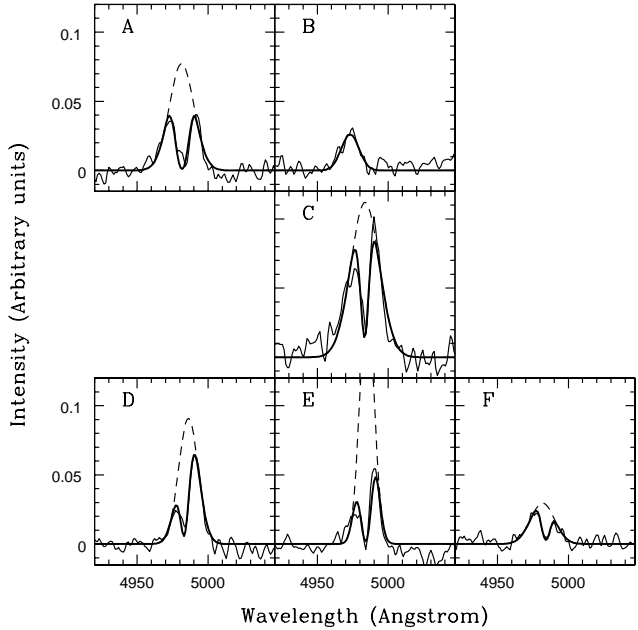,width=3.5in}}
\caption{\small Ly$\alpha$ emission line profiles across the galaxy LAB-2 at z=3.09. A profile is shown for each region labelled in Fig. 2. Thick solid lines are fits to a model where the intrinsic Ly$\alpha$ emission (dashed lines) is partially absorbed by foreground HI. The measured HI column densities are $\sim 10^{19}$~cm$^{-2}$, although the modest spectral resolution implies $1 \sigma$ uncertainties of at least a factor of $\sim 10$, so the column density may not be uniform across the galaxy. The widths of the absorbers (expressed as the gaussian b-parameter in the Voigt profile) are $\sim 20$~km~s$^{-1}$, which for thermal broadening implies a kinetic temperature $\sim 10^{4}$~K (as expected on photoionization grounds and consistent with the lack of bulk motion across the shell). Note, however, that similar Ly$\alpha$ profiles can result from a Ly$\alpha$ source in an HI slab of much higher optical depth$^{19}$; the twin peaks then arise because the escaping photons scatter into the line wings where the optical depth is lower. This would not, however, explain why the apparent absorption feature is constant in wavelength while the fitted emission velocity varies by almost 300~km~s$^{-1}$.}\end{figure}

\begin{figure}[t]
\centerline{\psfig{file=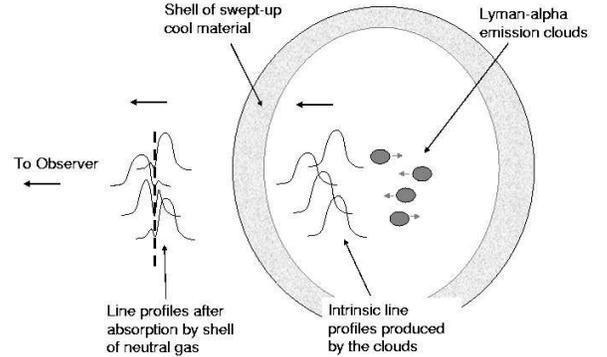,width=3.5in}}
\caption{\small An illustration of our model for the Ly$\alpha$ emission and absorption in the redshift z=3.09 galaxy LAB-2. The Ly$\alpha$ emission is produced by gas with a velocity dispersion $\sim 1000$~km~s$^{-1}$ and bulk motions $\sim 290$~km~s$^{-1}$. This emission is then absorbed/scattered by HI in a foreground shell. The shell's velocity coherence across the source suggests that it is over 100~kpc from galaxy, having been swept up from the surrounding medium by a starburst-driven outflow several 10$^{8}$~yr earlier. The possibility that the absorption is not directly associated with LAB-2, but part of a larger foreground absorber which just happens to cover it, is unlikely since this would imply an overall size much larger than absorbers of comparable N$_{\rm{HI}}$ in quasar sightlines$^{20}$: the LAB-2 absorber column density is between that of a Lyman-limit system (diameter $\sim 80$~kpc) and that of a damped Ly$\alpha$ absorber ($\sim 30$~kpc). Such sizes are consistent with recent modelling$^{21}$ and cosmological simulations$^{22}$. Moreover, it has been shown$^{23}$ that all damped Ly$\alpha$ absorbers at redshift $z \sim 3$ with N$_{\rm{HI}} > 2 \times 10^{20}$~cm$^{-2}$ can be explained by LBG galactic winds, if they are driven out to $\sim 27$~kpc.}
\end{figure}

        We now consider the cooling of the shocked gas. Gas behind a shock of speed v$_{\rm{shock}}$ is heated to $T = 1.4 \times 10^{7}$ (v$_{\rm{shock}}$/1000~km~s$^{-1}$)$^{2}$~K, from which the radiative cooling time is $t_{\rm{cool}}=1.5 kT/4n_{\rm{halo}} \Lambda$, where $\Lambda$ is the cooling function$^{16}$ ($>5 \times 10^{-23}$~erg~cm$^{3}$~s$^{-1}$ for $10^{6-7}$~K and solar abundances) and n$_{\rm{halo}}$ is the surrounding gas density (the factor 4 reflects the density increase behind a strong shock). Together, these imply that $t_{\rm{cool}} < 4.6 \times 10^{5}$~yr~(v$_{\rm{shock}}$/1000~km~s$^{-1}$)$^{2}$/(n$_{\rm{halo}}$/cm$^{-3}$). Moving out from the galaxy into the IGM at 100~kpc, the density n$_{\rm{halo}}$ will range from a few $10^{-3}$ to a few $10^{-4}$~cm$^{-3}$, so for shock velocities of several hundred kilometres per second, the shocked gas can cool within the expansion timescale of a few 10$^{8}$~yr. The gas is expected$^{17}$ to cool to $\sim 10^{4}$~K and come into photoionization equilibrium with the metagalactic ultraviolet radiation field, with an H I fraction $\sim 0.1-0.001$. Indeed, for the above n$_{\rm{halo}}$ values, the total column density swept up out to 100~kpc could be up to 10$^{21}$~cm$^{-2}$, compared with the measured N$_{\rm{HI}} \simeq 10^{19}$~cm$^{-2}$. The implied mass of the shell, if it covers an area 100~kpc square, could be almost 10$^{11}$~M$_{\rm{\odot}}$. For a velocity of 250~km~s$^{-1}$ the kinetic energy of such a shell is $6.3 \times 10^{58}$~erg, which we can compare with the energetics of supernovae: the LBG associated with LAB-2 (M14) has a near-infrared K$_{\rm{S}}$-band magnitude$^{4}$ of 21.7, which implies (from galaxy evolution models$^{18}$) a baryonic mass of $\sim 10^{11}$~M$_{\rm{\odot}}$, mostly in stars. For a normal stellar initial mass function, supernovae provide $\sim 10^{49}$~erg per solar mass of stars$^{1}$, so $\sim 10^{60}$~erg is available and the superwind model appears energetically feasible.

        We thus have a consistent picture in which the Ly$\alpha$ absorber is a highly ionized shell of gas swept up from the IGM over several $10^{8}$~yr since the end of the starburst, having slowed to within a few hundred kilometres per second of the galaxy's systemic velocity. This demonstrates that superwinds are a galaxy-wide phenomenon, whose impact is not confined to small-scale regions of current star formation, as previous observations might have suggested. This underscores their importance as a primary feedback agent in galaxy formation.


\vspace{0.5cm} 

\smallskip \smallskip
 
\noindent {\small {\bf Acknowledgements.} 
We thank the SAURON team for supporting this programme, together with E. Emsellem, E. Jourdeuil and ING staff for support on La Palma. The construction of SAURON was financed by contributions from ASTRON/NWO, the Institut des Sciences de l'Univers, the Universit\'e Claude Bernard Lyon-I, the universities of Durham and Leiden, The British Council, PPARC, and NOVA. R.J.W. and R.G.B. acknowledge support from PPARC, and R.G.B. also thanks the Leverhulme Foundation. J.G. is supported by the Euro3D research training network.}

\medskip
\noindent {\small {\bf Competing interests statement.} The authors
  declare that they have no competing financial interests.}

\medskip
\noindent {\small {\bf Correspondence} and requests for materials should be
  addressed to RJW (e-mail: r.j.wilman@durham.ac.uk).} 

\end{document}